\documentclass[letterpaper,iop,apj,numberedappendix,twocolappendix]{emulateapj}
\usepackage{thumbpdf} 
\usepackage{amsmath,amssymb} 
\usepackage{graphicx,mathptmx} 
\usepackage{natbib} 
\usepackage[usenames]{color} 
\usepackage{ifpdf}  
\usepackage[protrusion=true,expansion=false]{microtype} 
\usepackage{xcolor}   
\usepackage{refcount} 
\usepackage{textcase} 
\usepackage{epstopdf} 

\ifpdf
\pdfoutput=1 
\else

\fi

\newcommand{\angstrom}{\mbox{\normalfont\AA}}

\newcommand{\beq}{
\begin{equation}
}
\newcommand{\eeq}{
\end{equation}
}
\newcommand{\beqa}{
\begin{eqnarray}
}
\newcommand{\eeqa}{
\end{eqnarray}
}

\newcommand{\kgfigbeg}[1]{
\begin{figure}
\hypertarget{#1}{}%
}
\newcommand{\kgfigend}[2]{
\label{f:#1}
\end{figure}
\bookmarksetup{color=[rgb]{0.54,0,0}}
\bookmark[rellevel=1,keeplevel,dest=#1]{Fig \ref*{f:#1}: {#2}}
\bookmarksetup{color=black} 
}

\newcommand{\kgfigstarbeg}[1]{
\begin{figure*}
\hypertarget{#1}{}%
}
\newcommand{\kgfigstarend}[2]{
\label{f:#1}
\end{figure*}
\bookmarksetup{color=[rgb]{0.54,0,0}}
\bookmark[rellevel=1,keeplevel,dest=#1]{Fig \ref*{f:#1}: {#2}}
\bookmarksetup{color=black} 
}

\newcommand{\kgtabbeg}[1]{
\begin{deluxetable}{#1}
}
\newcommand{\kgtabend}[2]{
\label{t:#1}
\end{deluxetable}
\bookmarksetup{color=[rgb]{0,0,0.54}}
\bookmark[
rellevel=1,
keeplevel,
dest=table.\getrefnumber{t:#1}
]{Table \ref*{t:#1}: #2}
\bookmarksetup{color=[rgb]{0,0,0}}
}

\newcommand{\kgtabstarbeg}[1]{
\begin{deluxetable*}{#1}
}
\newcommand{\kgtabstarend}[2]{
\label{t:#1}
\end{deluxetable*}
\bookmarksetup{color=[rgb]{0,0,0.54}}
\bookmark[
rellevel=1,
keeplevel,
dest=table.\getrefnumber{t:#1}
]{Table \ref*{t:#1}: #2}
\bookmarksetup{color=[rgb]{0,0,0}}
}

\providecommand{\ion}[2]{#1$\;$\textsmaller{\@Roman{#2}}}

\def\spose#1{\hbox to 0pt{#1\hss}}
\newcommand{\lta}{\mathrel{\spose{\lower 3pt\hbox{$\mathchar"218$}}
      \raise 2.0pt\hbox{$\mathchar"13C$}}}
\newcommand{\gta}{\mathrel{\spose{\lower 3pt\hbox{$\mathchar"218$}}
      \raise 2.0pt\hbox{$\mathchar"13E$}}}
\def\simlt{\mathrel{\rlap{\lower 3pt\hbox{$\sim$}}\raise 2.0pt\hbox{$<$}}}
\def\simgt{\mathrel{\rlap{\lower 3pt\hbox{$\sim$}} \raise 2.0pt\hbox{$>$}}}

\definecolor{KayhanCiteColor}{rgb}{0,0.08,0.35}
\definecolor{KayhanURLColor}{rgb}{0,0.08,0.35}
\definecolor{KayhanLinkColor}{rgb}{0,0.08,0.35}
\definecolor{KayhanPageColor}{rgb}{0,0.08,0.35}
\definecolor{medred}{rgb}{0.75,0.0,0.0}

\shorttitle{A Multi-wavelength Analysis of PSO J334}
\shortauthors{Foord et al.}

\makeatletter
\ifx\undefined\hyperref
\usepackage[pdftitle={A Multi-Wavelength Analysis of Binary-AGN Candidate PSO J334.2028+01.4075},pdfauthor={Adi Foord},pdfsubject={Astrophysics},pdfkeywords={galaxies:nuclei --- galaxies:active ---- galaxies:jets ---- accretion, accretion disk --- black hole physics},letterpaper,linktocpage,colorlinks=true,linkcolor={KayhanLinkColor},urlcolor={KayhanURLColor},citecolor={KayhanCiteColor},hyperfootnotes=false]{hyperref}
\else\relax\fi
\makeatother
\usepackage[figure,figure*]{hypcap} 
\usepackage{atveryend} 
\usepackage[
  open,
  openlevel=3,
  atend
]{bookmark}[2010/04/08]  
\begin{document}

\label{firstpage}
 
\title{A Multi-wavelength Analysis of Binary-AGN Candidate PSO J334.2028+01.4075}

\author{Adi Foord\altaffilmark{1}}
\author{Kayhan G\"{u}ltekin\altaffilmark{1}}
\author{Mark Reynolds\altaffilmark{1}}
\author{Megan Ayers\altaffilmark{2,3}}
\author{Tingting Liu\altaffilmark{4}}
\author{Suvi Gezari\altaffilmark{4}}
\author{Jessie Runnoe\altaffilmark{1}}
\affil{\altaffilmark{1}Department of Astronomy, University of Michigan, 1085 S. Univ., Ann Arbor, MI 48103, USA;
\href{mailto:foord@umich.edu}{foord@umich.edu}.}

\affil{\altaffilmark{2}Department of Physics, Lewis \& Clark College, 0615 SW Palatine Hill Rd, Portland, OR 97219, USA}

\affil{\altaffilmark{3}Maria Mitchell Observatory, 4 Vestal Street, Nantucket, MA 02554, USA}

\affil{\altaffilmark{4}Department of Astronomy, University of Maryland, College Park, MD 20742, USA}

\begin{abstract}
\hypertarget{abstract}{} 
We present analysis of the first \emph{Chandra} observation of PSO J334.2028+01.4075 (PSO J334), targeted as a binary-AGN candidate based on periodic variations of the optical flux.  With no prior targeted X-ray coverage for PSO J334, our new 40 ksec \emph{Chandra} observation allows for the opportunity to differentiate between a single or binary-AGN system, and if a binary, can characterize the mode of accretion.  Simulations show that the two expected accretion disk morphologies for binary-AGN systems are (i) a ``cavity," where the inner region of the accretion disk is mostly empty and emission is truncated blueward of the wavelength associated with the temperature of the innermost ring, or (ii) ``minidisks", where there is substantial accretion from the cirum-binary disk onto one or both of the members of the binary, each with their own shock-heated thin-disk accretion system. We find the X-ray emission to be well-fit with an absorbed power-law, incompatible with the simple cavity scenario. Further, we construct an SED of PSO J334 by combining radio through X-ray observations and find that the SED agrees well with that of a normal AGN, most likely incompatible with the minidisk scenario.  Other analyses, such as locating the quasar on IR color-color diagrams and analyzing the quasar mass predicted by the fundamental plane of black hole activity, further highlight the similarity of PSO J334 with respect to normal AGN.  On the multi-wavelength fronts we investigated, we find no evidence supporting PSO J334 as a binary-AGN system, though our analysis remains insensitive to some binary configurations.

\bookmark[ rellevel=1, keeplevel,
dest=abstract
]{Abstract}
\end{abstract}

\section{Introduction}  
\label{intro}
Classical heirarchical galaxy evolution predicts galaxies to merge (e.g., \citealt{WhiteandReese1978}), allowing any central supermassive black holes (SMBH)  to assemble into binary active galactic nuclei (AGN) systems \citep{Volonteri2003}.  The galaxy merger can serve as one possible avenue of growth for the central black holes, and it is expected to end with the coalescence of the two black holes as a result of the emission of gravitational waves (see \citealt{Begelman1980}).  The new SMBH will have a different mass and spin \citep{Rezzolla2008} and can receive a gravitational recoil large enough to kick it out of the merged host galaxy (e.g., \citealt{Volonteri2008,Baker2008, Merritt2004, Lousto2013}).  As powerful sources of gravitational waves and a likely influence on the BH occupation fraction of galaxies, binary SMBHs are important systems to study.

The process of black hole merging can be broken into distinct phases. Here, we reorganize the original phases presented in \cite{Begelman1980} to emphasize details important to our analysis: (1) the galaxy merger phase, where the two central black holes sink to the center and a bound black hole pair forms with semimajor axis $a$; (2) the final parsec phase, where the black hole binary system may or may not stall at a separation of $0.1 \lta a \lta 1.0$ pc as a result of ejecting all the stars in its loss cone.  This phase is referred to as the ``final parsec problem" \citep{Milosavljevic2003, MilosavljevicMerritt2003} and many potential solutions have been theorized (e.g., \citealt{Yu2002,Escala2005,Berczik2006,Mayer2007,Dotti2007,Berentzen2009,Cuadra2009,Lodato2009,Khan2013}); (3) the circum-binary accretion phase, where circum-binary accretion is expected as a result of the typical accretion disk size being larger than the binary separation $a$ (see \citealt{Milosavljevic2005}); (4) the gravitational wave phase, where the binary is sufficiently hardened and gravitational waves carry energy from the system until the binary merges (see \citealt{Peters1964}); and (5) the post-merger phase, where the new black hole has a different mass and spin \citep{Rezzolla2008}, possibly leading to a recoil large enough to displace or eject the black hole from the galaxy center \citep{Volonteri2008,Baker2008, Merritt2004, Lousto2013}.

``Dual-AGN" are usually defined as a pair of AGN with kilo-parsec scale separations (i.e., phase 1; see \citealt{Comerford2009}), while a ``binary-AGN" is a pair of BHs that are gravitationally bound with typical separations $a < 100$ pc (i.e., phase 2 \& 3; see \citealt{Bansal2017} for the first resolved binary-AGN candidate with a separation of $\sim7.3$ pc).  Such a system becomes impossible to resolve with \emph{Chandra} beyond a distance of $\sim$4 Mpc.  For example, the closest dual-AGN candidate identified using two resolved point sources with \emph{Chandra} is NGC 3393 \citep{Fabbiano2011} with a projected separation of $\sim$150 pc ($\sim$0\farcs6).  However, this source has been contested as potentially spurious (e.g.,  \citealt{Koss2015}). 

Thus many indirect detection techniques have been developed to search for signs of binary-AGN.  One such method involves looking for periodic variability in the optical flux via time-domain observations, a possible result of accretion via a circum-binary disk (e.g., \citealt{D'Orazio2013, Farris2015a, Farris2015b}). Perhaps the strongest case of a candidate binary-AGN, OJ 287, was identified by its variable luminosity and has exhibited regular optical outbursts with $\sim$12 year period \citep{Lehto1996}.  However, OJ 287 is not the typical binary system, as the fluctuations in its lightcurve have been modeled as the secondary SMBH periodically intercepting the primary SMBH's accretion disk (\citealt{Valtonen2008} and references therein).  Such a model can result from a configuration where there is a considerable misalignment between the orbital plane of the secondary SMBH and the accretion-disk plane of the primary SMBH.

Other quasars have been identified as binary-AGN candidates via time-domain techniques, such as PG 1302$-$102 \citep{Graham2015} and PSO J334.2028+01.4075 (hereafter PSO J334; \citealt{liu2015}). PG 1302$-$102 was identified as a binary-AGN based on the periodic variability of the optical flux on an observed timescale of $\sim$1,884 days, corresponding to a separation of $a\sim0.01$ pc \citep{Graham2015}.  Further, it was argued that the variability of the light curve could be explained by relativistic Doppler boosting from an unequal-mass binary \citep{D'Orazio2015}. Similarly, PSO J334, at a redshift of $z=2.06$, was identified as a potential binary system based on periodic variation of the optical flux on an observed timescale of $\sim$542 days, corresponding to a separation $a\sim0.006$ pc \citep{liu2015}. However, recently \cite{Vaughan2016} have shown that the data presented on PG 1302-102 and PSO J334 are not strong enough to support the model of a binary-AGN system. Specifically, they find that sinusoidal variations are difficult to distinguish from a stochastic (``red noise") process when the number of cycles is $\le$ 2, and that at least $\sim$5 cycles are needed to confirm a true periodic trend in lightcurves.  In response, \cite{liu2016} tested the persistence of PSO J334's periodic lightcurve fluctuations using an extended baseline analysis composed of both archival and new data.  This new analysis disfavors a simple sinusoidal model for PSO J334 over a baseline of ∼5 cycles. Yet, the true nature of PSO J334 remains in question.  Recent Karl Jansky Very Large Array (VLA) and Very Long Baseline Array (VLBA) coverage presented in \cite{mooley2017} further supports that PSO J334 is a binary black hole system, as the quasar was found to be lobe-dominated with a twisted radio structure, a possible result of a precessing jet.  

Perhaps the best way to discern between the binary and single-AGN models is to use multi-wavelength observations.  With no prior targeted X-ray coverage for PSO J334, our new 40 ksec \emph{Chandra} observation allows for a complete multi-wavelength description of the quasar.  Specifically, combining archival data with our new observations may enable us to differentiate between a single or binary-AGN system, and if a binary, can possibly characterize the mode of accretion. If PSO J334 is a binary system with separation $a = 28R_{S}$ (where $R_{S}={2GM}/{c^{2}}$ is the Schwarzchild radius for a BH with mass $M$; \citealt{liu2016}) we expect that the binary is well into the gravitational-wave dominated regime, where circum-binary accretion is likely. The mode of circum-binary accretion will depend on the current state of the system. For example, if the specific angular momentum of the streams is small compared to the specific angular momentum at the innermost stable circular orbit (ISCO), the streams will flow directly into the SMBHs (see \citealt{Gultekin&Miller2012}, \citealt{Tanaka&Haiman2013}, \citealt{Tanaka2013}, \citealt{Gold2013}, \citealt{Roedig2014}). However, this scenario is expected for SMBHs with very small separations.  For all other binary systems, accretion disks form around each SMBH (``minidisks''), extending to a tidal truncation radius that is expected to be less than $\sim$ $a$/2 (\citealt{Paczynski1977}, also see \citealt{Roedig2014} for the dependency of the truncation radius on mass ratio $q$). 

For further-evolved binaries, the timescale to fully accrete the minidisks can be smaller than the gravitational-wave timescale; in this scenario the mini-disks will be drained before the two SMBHs merge.  Here we may expect a cavity between the circum-binary disk and the SMBHs, as the inner regions of the accretion disk are mostly void of gas. The cavity model and the minidisk model are expected to manifest differently in the observational data. For example, one can look at the X-ray spectrum to search for the presence of streams from the circum-binary disk accreting onto the minidisk (e.g., \citealt{Roedig2014, Farris2015a, Farris2015b}).  Further, the radio--X-ray AGN spectral energy distribution (SED) can be used to search for abnormalities, such as ``notches" in the SED expected from minidisks (however, see \citealt{Leighly2016} for a critical perspective on possible ``notches" in the SED of Mrk 231), or the presence of a cavity (see \citealt{Milosavljevic2005}).  Here we present a multi-wavelength analysis of PSO J334, discovered in the FIRST Bright Quasar Survey \citep{Becker2001}, in order to infer its true accretion nature. In the following analysis we assume a standard $\Lambda$CDM cosmology of $\Omega_{\Lambda} = 0.7$, $\Omega_{M} = 0.3$, and $H_{0} = 70$ km s$^{-1}$.

\section{X-ray Data Analysis}
\label{analysis}
We targeted PSO J334 in Cycle 17 (Proposal ID:17700741, PI: G\"{u}ltekin). The quasar was placed on the back illuminated S3 chip of the Advanced CCD Imaging Spectrometer (ACIS) detector, with an exposure time of 40 ks.  The exposure time was chosen in order to achieve a 3$\sigma$ point-like source detection, based on an upper-limit calculated from a previous \textit{XMM-Newton} slew survey observation where no emission consistent with the position of PSO J334 was detected, and assuming emission from an active black hole with $L$ $\ge$ 10$^{-3}$ $L_{\mathrm{Edd}}$ with mass log$(M/M_{\odot})=9.1$ \citep{liu2016,mooley2017}. Our 40 ks exposure is not sensitive to the $\sim$542 day optical period found by \cite{liu2016}. Further, our observation is not sensitive to any quasi-periodic signal associated with the decay time from gravitational waves.  

We follow a similar data reduction as described in previous X-ray studies analyzing active fractions \citep{Gallo2008,Gallo2010,Miller2012a,Miller2012b,Miller2015,Plotkin2014,Foord2017}, using the Chandra Interactive Analysis of Observations ({\tt CIAO}) v4.8.  We first correct for astrometry, cross-matching the \emph{Chandra}-detected point-like sources with the Sloan Digital Sky Survey Data Release 9 (SDSS DR9) catalog. The \emph{Chandra} sources used for cross-matching are detected by running {\tt wavdetect} on the reprocessed level 2 event file.  We require each matched pair to be less than 2$\arcsec$ from one another and have a minimum of 3 matches. Our final astrometrically-corrected image has a shift less than 0\farcs5.  We then correct for background flaring by removing intervals where the background rate was found to be 3$\sigma$ above the mean level, resulting in the removal of a 197 second interval.  We then rerun {\tt wavdetect} on filtered 0.5 to 7 keV data to generate a list of X-ray point sources. We use wavelets of scales 1, 1.5, and 2.0 pixels using a 1.5 keV exposure map, and set the detection threshold significance to 10$^{-6}$ (corresponding to one false detection over the entire S3 chip). We identify the quasar as an X-ray point source $\sim0\farcs4$ from the nominal, SDSS-listed optical center (2$\arcsec$ corresponds to 95\% of the encircled energy radius at 1.5 keV for ACIS). 

All errors evaluated in this paper are done at the 95\% confidence level and error bars quoted in the following section are calculated with Monte Carlo Markov Chains via the XSPEC tool {\tt chain}.

\subsection{Spectral Fitting}
The quasar's net count rate and flux value are determined using XSPEC, version 12.9.0 \citep{Arnaud1996}. Counts are extracted from a circular region with radius of 2$\arcsec$ centered on the X-ray source center, using a source-free annulus with inner radius of 20$\arcsec$ and outer radius of 30$\arcsec$ for the background extraction.  We fit the spectrum between 0.3 and 7 keV with an absorbed red-shifted power-law ({\tt phabs*zphabs*zpow}; hereafter Model 1) where we fix the Galactic hydrogen column density (the photoelectric absorption component {\tt phabs}) to a value\footnote{We evaluate the neutral hydrogen column density using values from the Leiden/Argentine/Bonn (LAB) Survey of Galactic HI \citep{Kalberla2005} via WebPIMMS.} of 3.5 $\times$ 10$^{20}$ cm$^{-2}$.  As a result of being in the low-count regime, we implement the Cash statistic ({\tt cstat}; \citealt{Cash1979}) and a minimum of 1 count per bin in order to best assess the quality of our model fits.  We find the best-fit parameters intrinsic $N_{H} = 0.91^{+4.84}_{-0.89} \times 10^{22}$ cm$^{-2}$  and $\Gamma = 2.02^{+0.83}_{-0.39}$, with an observed 2--10 keV flux of $3.20^{+0.9}_{-1.1} \times$ 10$^{-14}$ erg cm$^{-2}$ s$^{-1}$, or rest-frame 2--10 keV luminosity of $9.40^{+1.4}_{-1.1} \times$ 10$^{44}$ erg s$^{-1}$ at $z=2.06$ (assuming isotropic emission).  $K$-corrections are not applied to the \emph{Chandra} data, as we directly measure the flux density from the spectrum.  In Figure~\ref{fig:1}, we show the X-ray spectrum of PSO J334 along with the best-fit XSPEC model.  

We add a line component to the \emph{Chandra} spectrum ({\tt phabs*zphabs*(zpow+zgaus)}) to investigate the presence of an Fe K-$\alpha$ line (seen as a slight excess compared to the model at observed-frame 2 keV in Fig.~\ref{fig:1}).  Allowing the line energy to vary, a gaussian component is best-fit at rest-frame $6.2^{+1.2}_{-4.4}$ keV.  We find that the addition of this Fe K-$\alpha$ line is not statistically significant (as is evident from the uncertainties on the line energy) most likely a result of the spectrum only having 196 counts. Further, the placement of the emission line is near a strong drop in effective area of the ACIS-S3 chip, which complicates the study of a potential line.  While the Fe K-$\alpha$ line is not statistically significant, we fix the rest-frame energy to 6.4 keV and the width of the line $\sigma = 0$ keV, and calculate an equivalent width (EW) of 0.22 keV with a $3\sigma$ upper limit of 0.55 keV.

We test for the presence of two accretion disks by fitting a broken power-law to the spectrum ({\tt phabs*zphabs*bknpo}; Model 2). Such a spectrum may originate from a binary system where both SMBHs are accreting with their own minidisk.  We model the spectrum with a broken power-law in order to avoid the degeneracies that exist in parameter-space for a double power-law model.  To properly compare Model 2 to a single disk system, we also fit the spectrum with a broken power-law but tie the two photon index values to one-another (i.e, $\Gamma_{1} = \Gamma_{2}$; Model 3).  We note this approach produces best-fit parameters consistent with best-fit parameters of Model 1. We conduct an F-test to investigate the significance of Model 2 with respect to Model 3, and find an f-value of 0.25 (with probability value $p = 0.78$). At a 95$\%$ confidence level, we conclude that the spectrum does not need an additional photon index to explain its shape.  We note that these results are not necessarily indicative of a single-AGN system, as it is difficult to disentangle two power-laws without predominant spectral features, such as Fe K-$\alpha$ lines at different velocities (a non-negligible scenario, see \citealt{Eracleous2012,Popovic2012,Jovanovic2016,Simic2016} for more details of emission lines in binary-AGN).

We calculate the X-ray hardness ratio (HR) of PSO J334, defined as (H-S)/(H+S) where S and H are the soft ($0.5$--$2.0$ keV) and hard ($2.0$--$8.0$ keV) X-ray band net counts detected by \emph{Chandra}. The HR is found to be $\sim$-0.40, consistent with the expected value for an AGN with $N_{H}\simeq10^{22}$ cm$^{-2}$ and $z\simeq2.06$ (e.g. \citealt{Wang2004}).

\kgfigbeg{fig:1}
\centering
\includegraphics[width=8.5cm]{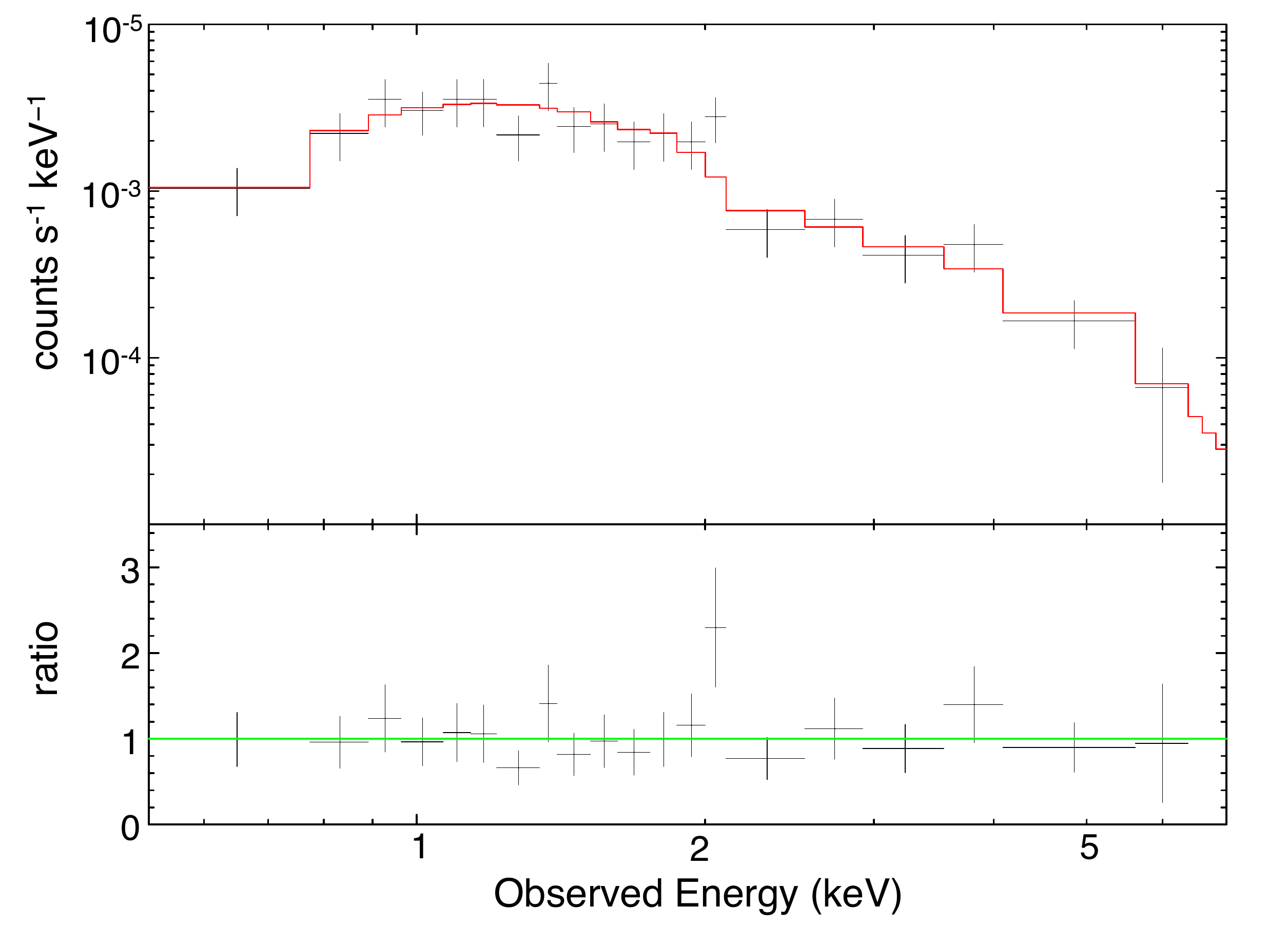}
\caption{\emph{Top:} The observed 0.3 -- 7.0 keV \emph{Chandra} spectrum of PSO J334 is shown in black, where the data have been folded through the instrument response.  We fit the spectrum with the model {\tt phabs*zphabs*zpow}, fixing the Galactic absorption and redshift parameters at $N_{H} = 3.5 \times$ 10$^{20}$ cm$^{-2}$ and $z=2.06$. The best-fit model is shown in red, where intrinsic $N_{H} = 0.91^{+4.84}_{-0.89} \times 10^{22}$ cm$^{-2}$  and $\Gamma = 2.02^{+0.83}_{-0.39}$.  We calculate an observed 2--10 keV flux of $3.20^{+0.9}_{-1.1} \times$ 10$^{-14}$ erg cm$^{-2}$ s$^{-1}$, or rest-frame 2--10 keV luminosity of $9.40^{+1.4}_{-1.1} \times$ 10$^{44}$ erg s$^{-1}$ at $z=2.06$ (assuming isotropic emission).  All errors are evaluated at the 95\% confidence level. \emph{Bottom:} Ratio of the data to the continuum model for PSO J334.  We find that at a 95$\%$ confidence level, the spectrum does not need an additional power-law to explain the data.  The spectrum has been rebinned for plotting purposes.}
\label{fig:1}
\kgfigend{fig:1}{X-ray spectrum}

\section{The Spectral Energy Distribution}
\label{sec:SED}
In the following section we construct a spectral energy distribution of PSO J334, by combining radio to X-ray observations, and compare it to standard non-blazar AGN SEDs presented in \cite{Shang2011}.  For all $K$-corrections, we adopt the $K(z)$ relation as presented in \cite{Richards2006}.

\subsection{Radio}
PSO J334 has archival spectroscopy from the FIRST Bright Quasar Survery (FBQS), and has recently been re-observed by the VLA as part of the Caltech-NRAO Stripe 82 Survey \citep{Mooley2016}. Further, \cite{mooley2017} present VLBA observations of PSO J334 at 7.40, 8.67, and 15.37 GHz, respectively. Two VLBA components are resolved\,---\,a South East and a North West component.  Both the compactness and the inverted radio spectrum of the South East component (possibly due to synchrotron self-absorption) suggest that it is the ``core" from which the North West component has been ejected.  For the purposes of constructing the SED, we use the VLBA integrated flux density values for the South East component (see table 3 in \citealt{mooley2017}).  $K$-corrections for the radio data points are implemented assuming a spectral index $\alpha = -0.3$ with a dispersion 0.2 (where $F_{\nu} \propto \nu^{-\alpha}$).  The value for $\alpha$ is taken from \cite{Stocke1992}, and represents the radio slope of the average quasi-stellar object (QSO) SED, which includes contributions from both the core and lobes.

\subsection{Infrared}
For the infrared regime we use archival \emph{Wide-field Infrared Survey Explorer} (\emph{WISE}) from the AllWISE Source Catalog \citep{ALLWISE2013}, as well as stacked archival $J$- and $K$-band data from the UKIRT InfraRed Deep Sky Survey (UKIDSS; \citealt{Lawrence2007}). For the \emph{WISE} data we use observations taken in the W1, W2, and W3 bands, where the quasar was detected with a SNR $>$ 3.0.  We correct for Galactic extinction using the dust map from \cite{Schlafly&Fink2011}, where $E(B-V) = 0.0401$. $K$-corrections are applied assuming a spectral index $\alpha = -1.0$ with a dispersion of 0.2, calculated from the average 1--10 $\mu$m spectral index for the AGN sample presented in \cite{Shang2011}.

\subsection{Optical}
PSO J334 has archival $g$-, $r$-, $i$-, and $z$-band data from the Pan-STARRS1 Medium Deep Survey (PS1 MDS; \citealt{Kaiser2010}), $V$-band data from the Catalina Real-time Transient Survey (CRTS; \citealt{Drake2009}), and archival $u$-, $g$-, $r$-, $i$-, and $z$-band data from the Sloan Digital Sky Suvey (SDSS).  For the purposes of constructing our SED, we use magnitudes extracted from deep stacked images in the $g$-, $r$-, $i$-, and $z$-bands from PS1 MDS \citep{liu2015, liu2016}.  Similar to the IR regime, we correct for Galactic extinction using the dust map from \cite{Schlafly&Fink2011}. For the PS1 MDS $K$-corrections, we follow \cite{liu2015} and assume a spectral index $\alpha = -0.5$ with a dispersion of 0.3 (also see \citealt{Elvis1994,Vandenberk2001,Ivezic2001}).

\subsection{Ultraviolet}
PSO J334 has $u$-band data from the Canada-France-Hawaii Telescope (CFHT; \citealt{Heinis2016a,Heinis2016b,liu2016}) and archival \emph{Galaxy Evolution Explorer} (\emph{GALEX}) data in both the FUV and NUV bands.  Further, the quasar has \emph{GALEX} Time Domain Survey data in the NUV, taken to analyze possible periodic variations in the UV lightcurve \citep{Gezari2013, liu2015, liu2016}.  For our SED we use magnitudes derived from deep stacked $u$-band CFHT data and the archival FUV and NUV \emph{GALEX} data. To account for Galatic absorption in the CFHT data we use the dust map from \cite{Schlafly&Fink2011}, while for the \emph{GALEX} observations we use FUV and NUV extinction values listed in \cite{Yuan2013}. As all data points are shortward of the Ly$\alpha$ emission line, we apply $K$-corrections using the spectral index $\alpha = -1.57$ and a dispersion of 0.17 \citep{Telfer2002,Richards2006}. \\
%

\kgfigstarbeg{fig:2}
\centering
\includegraphics[width=16.8cm]{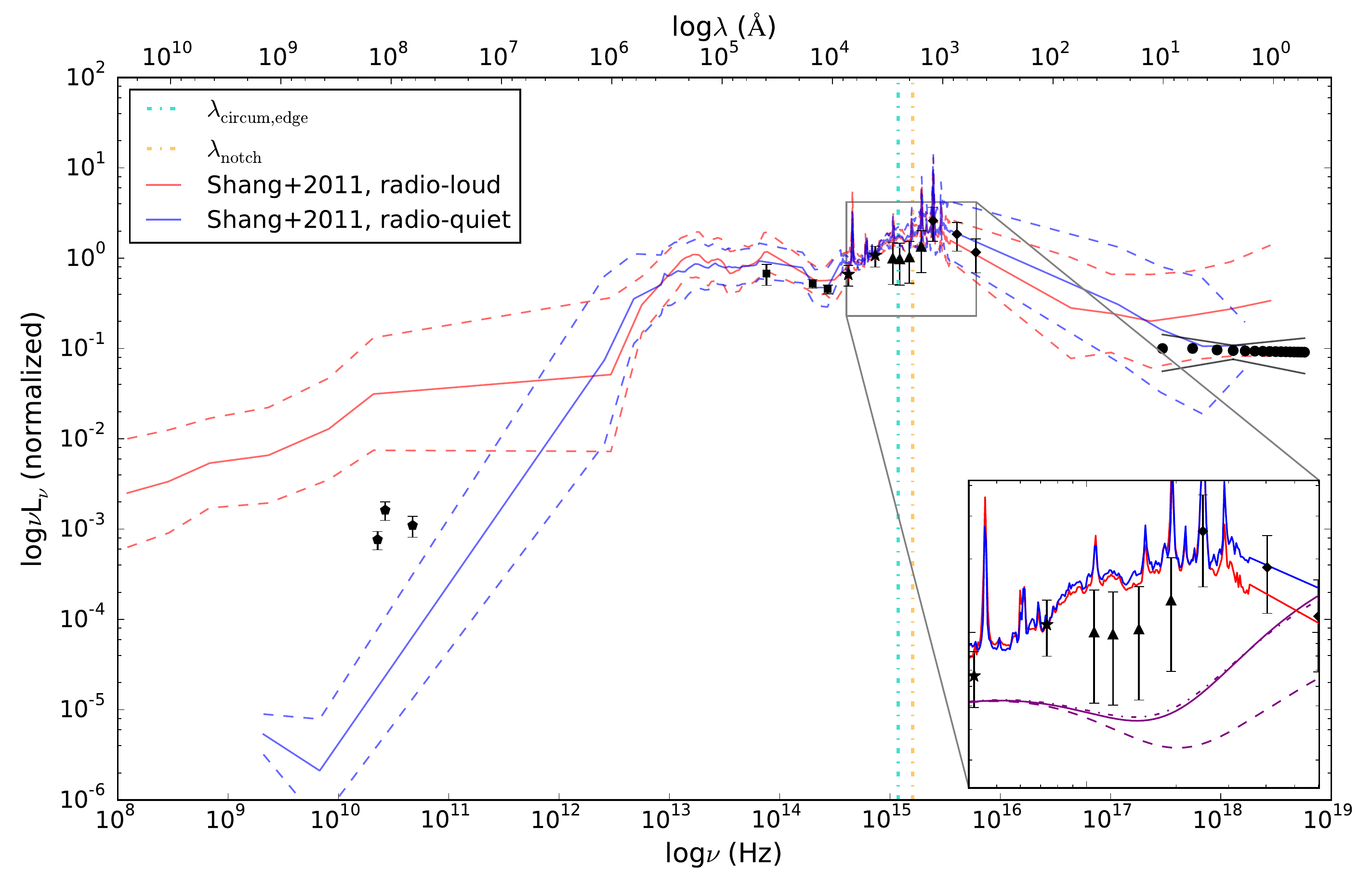}
\caption{Rest-frame SED of PSO J334.  Radio data are from the VLA (black hexagons), IR data are from \emph{WISE} (black-filled squares) and UKIDSS (black-filled stars), optical data are from Pan-STARRS1 (black-filled triangles) , UV data are from CFHT and \emph{GALEX} (black-filled diamonds), and X-ray data is from \emph{Chandra} (black-filled circles). Errors on data points are evaluated at the 95\% confidence level.  We overplot the composite non-blazar AGN SEDs presented in \cite{Shang2011}, for both radio-loud (red line) and radio-quiet (blue line) AGN (1$\sigma$ error bars are denoted by dashed lines). We normalize the flux density of our data to rest-frame $\lambda = 2000$ \angstrom. We indicate $\lambda_\mathrm{circum,edge} = 2500$ \angstrom, the wavelength that corresponds to the emission emitted at the inner-edge of a possible circum-binary accretion disk at $R_\mathrm{circum,edge} = 2a$ (assuming blackbody radiation), with a cyan dot-dashed line.  If PSO J334 were consistent with the cavity model, we do not expect much emission at wavelengths with energies higher than $\lambda_\mathrm{circum,edge}$. We also indicate the predicted center wavelength for a notch at $\lambda_\mathrm{notch} = 1900$ \angstrom, predicted to range between 500 \angstrom ~and 7000 \angstrom ~for a mass ratio $0.3 < q< 1.0$.  If PSO J334 were consistent with the minidisk model, we expect a dip in the thermal continuum in this region.  The inset shows how the SED is expected to change with the addition of a notch, where we use the analytical calculations derived in \cite{Roedig2014} to illustrate a notched SED with i) $q = 1.0$, $f_{1} = f_{2} = 0.5$ (purple solid curve), ii) $q = 0.3$, $f_{1} = 0.45$, $f_{2} = 0.55$ (purple dash-dot curve, and iii) $q = 0.1$, $f_{1} = 0.92$, $f_{2} = 0.08$ (purple dashed curve). The continuum for the notch is estimated by approximating the \cite{Shang2011} SED between 500 \angstrom ~and 7000 \angstrom~in log space with a straight line.  We note that although PSO J334 appears to be better aligned with the radio-quiet sample, the quasar is technically considered to be radio-loud with $R \sim 17$. }
\label{fig:2}
\kgfigstarend{fig:2}{Multi-wavelength SED}
%
\subsection{The Multi-wavelength SED}
In Fig.~\ref{fig:2} we present the rest-frame multi-wavelength SED of PSO J334 in $\nu$L$_{\nu}$ (erg s$^{-1}$) versus frequency (Hertz), assuming a luminosity distance $d_{L} = 16.1$ Gpc at $z = 2.06$.  We search for any abnormalities in the SED by comparing our data to the composite non-blazar AGN SEDs presented in \cite{Shang2011}. The composite SEDs are based on a sample of 85 optically bright non-blazar quasars, composed of 27 radio-quiet and 58 radio-loud quasars.  Most objects in the sample have quasi-simultaneous UV/optical data, while radio, IR, and X-ray data are obtained from either the literature or new observations. Both the radio-quiet and radio-loud composite SEDs are overplotted with PSO J334's data in Fig.~\ref{fig:2}.  We normalize the flux density of our data to rest-frame $\lambda = 2000$ \angstrom ~for comparison to the standard AGN SEDs. The luminosities for each rest-frame frequency are listed in Table~\ref{tab:SED}.

A simple comparison between the data and the \cite{Shang2011} SEDs shows a good agreement between PSO J334's emission and that of a radio-quiet AGN.  The classical definition of radio-quiet quasars are AGN that have $R = f$(5 GHz)/$f$(4400 \angstrom) $< 10$, where $f$(5 GHz) is the rest-frame flux density at 5 GHz and $f$(4400 \angstrom) is the rest-frame flux density at 4400 \angstrom~ \citep{Kellermann1989}. Above $R = 10$, AGN are usually classified as radio-loud. Results from \cite{Becker2001} identify PSO J334 as a radio-loud quasar, with $R$ $\sim$200. We use the new 7.40 GHz observations of the South East component of PSO J334 from \cite{mooley2017}, where the resolution has increased from the previous FBQS observations, and derive an $R$ value of $\sim17$ (similar to \citealt{Becker2001}, we assume the radio spectrum follows $f \propto \nu^{-0.5}$ to extrapolate the rest-frame 5 GHz flux density).  Although PSO J334 is technically defined as a radio-loud AGN, $R\sim17$ is decidedly between the radio-loud sample SED from \cite{Shang2011} (where the median $R\sim1600$) and the radio-quiet sample SED (where the median $R\sim0.3$).  Besides the radio regime, there is substantial overlap between PSO J334's SED and the \cite{Shang2011} data at all other wavelengths.
%

\begin{table}[t]
\begin{center}
\caption{Spectral Energy Density Values}
\label{tab:SED}
\small
\begin{tabular}{llccccc}
	\hline
	\hline
	\multicolumn{1}{c}{Filter or} & \multicolumn{1}{c}{Telescope or} & \multicolumn{1}{c}{log$\nu$} & \multicolumn{1}{c}{log$\nu$L$_{\nu}$}   \\
	\multicolumn{1}{c}{Detector} & \multicolumn{1}{c}{Survey} & \multicolumn{1}{c}{(log(Hz))} & \multicolumn{1}{c}{(log(erg s$^{-1}$))}  \\
	\multicolumn{1}{c}{(1)} & \multicolumn{1}{c}{(2)} & \multicolumn{1}{c}{(3)} & \multicolumn{1}{c}{(4)} \\
	\hline
	\dots	&	VLBA	&	10.35 		&	$-3.12\pm0.11$	\\
    \dots	&	VLBA	&	10.42		&	$-2.80\pm0.11$	\\
	\dots	&	VLBA	&	10.67		&	$-2.96\pm0.13$	\\
	W3	&	\emph{WISE}	&	13.88		&	$-0.02\pm0.13$	\\
	W2	&	\emph{WISE}	&	14.30		&	$-0.14\pm0.05$	\\
	W1	&	\emph{WISE}	&	14.44		&	$-0.20\pm0.05$	\\
	$J$	&	UKIDSS	&	14.62		&	$-0.03\pm0.13$	\\
	$K$	&	UKIDSS	&	14.87		&	$0.18\pm0.13$	\\
	$z$	&	PS1 MDS	&	15.03		&	$0.15\pm0.29$	\\
	$i$	&	PS1 MDS	&	15.18		&	$0.14\pm0.29$	\\
	$r$	&	PS1 MDS	&	15.09		&	$0.16\pm0.29$	\\
	$g$	&	PS1 MDS	&	15.28		&	$0.28\pm0.29$	\\
	$u$	&	CFHT	&	15.40		&	$0.56\pm0.23$	\\
	NUV &   \emph{GALEX} &  15.61   &   $0.21\pm0.22$ \\
    FUV	&	\emph{GALEX} &  15.78	&	$0.41\pm0.19$	\\
    ACIS-S3 & \emph{Chandra} & 17.68--18.38  & $-0.87^{+0.06}_{-0.05}$ \\
	\hline
\end{tabular}
\end{center}
\medskip
Note. -- Columns: (1) filter or detector; (2) telescope, denoted as the Very Large Baseline Array (VLBA); the \emph{Wide-field Infrared Survey Explorer} (\emph{WISE}); the UKIRT InfraRed Deep Sky Survey (UKIDSS); the Pan-STARRS1 Medium Deep Survey (PS1 MDS); the Canada-France-Hawaii Telescope (CFHT); the \emph{Galaxy Evolution Explorer} (\emph{GALEX}); and the \emph{Chandra X-ray Observatory} (3) rest-frame frequency assuming a redshift of $z = 2.06$, in units of Hz. The \emph{Chandra} frequency range corresponds to the rest-frame energy range of 2--10 keV; (4) extinction- and $K$-corrected luminosity assuming a luminosity distance $d_{L}=$16.1 Gpc, in units of erg s$^{-1}$.  Values have been normalized to the luminosity at rest-frame $\lambda = 2000$\angstrom.  Error bars are evaluated at the 95\% confidence level.  Please see Section~\ref{sec:SED} for details on extinction values and $K$-corrections applied.

\end{table}

\section{Results and Discussion}
Past studies of PSO J334 have revealed various results regarding the accretion mode of the quasar, as analyses in different wavelength regimes have suggested both single- and binary-AGN systems.  PSO J334 was first targeted as a binary-AGN candidate based on periodic variation of the optical flux \citep{liu2015}.  Assuming that the rest-frame period of the quasar variability traced the orbital period of the binary, a binary separation of $\sim 0.006$ pc was estimated.  However, recently \cite{Vaughan2016} have shown that sinusoidal variations are difficult to distinguish from a stochastic process when the number of cycles is $\le$ 2; they conclude that at least $\sim$5 cycles are needed to confirm a true periodic trend in lightcurves. Searching for sinusoidal variations in candidate binary-AGN lightcurves is complicated by the fact that binary-AGN are likely to show both periodic and stochastic variations.  This is a result of regular quasar variability overlapping with any modulation resulting form being in a binary system.  More complex analyses will be necessary for future studies of candidate binary-AGN lightcurves, as it is still relatively uncertain how to best model quasar noise power spectra.

\cite{liu2016} confirmed that the data disfavored a simple sinusoidal model using an extended baseline analyses composed of both new and archival data in optical and UV.  However, most recently \cite{mooley2017} have re-established the idea of PSO J334 being a binary-AGN system via new VLA and VLBA observations.  The central radio ``core" of the quasar was found to have an elongation position angle (PA) twisted by $\sim39^{\circ}$ with respect to the elongation PA measured by the VLA on kiloparsec scales.  Such twists have been modeled in 3C 207 as a result of a precessing jet, possibly due to the orbital motion associated with binary-AGN (see \citealt{Hough2013}).  However, if the jet axis is close to our line of sight, the PAs may appear amplified as projected onto the plane of the sky.  Such a scenario may be relevant to PSO J334, which has been identified as a Type I quasar via broadened CIV (1549 \angstrom) and Mg II (2798 \angstrom) lines \citep{liu2015,liu2016,mooley2017}.  

With no prior targeted X-ray coverage for PSO J334, our new 40 ksec \emph{Chandra} observation allows for a complete multi-wavelength description of the quasar.  Specifically, combining radio--X-ray observations enables us to differentiate between a single or binary-AGN system, and if a binary, can possibly characterize the mode of accretion.  Simulations show that the two most basic types of accretion disk morphologies for binary-AGN systems are a “cavity”, where the inner region of the accretion disk is mostly empty and emission is truncated blueward of the wavelength associated with the temperature of the innermost ring, or “minidisks”, where there is substantial accretion onto one or both of the members of the binary, each with their own shock-heated thin-disk accretion system. In the following section we will discuss the implications of our \emph{Chandra} observation and investigate the possibility of PSO J334 being a binary-AGN in terms of the cavity and mini-disk models.

\subsection{The Null Hypothesis: PSO J334 is a single-AGN}
We first consider analyses which are used to identify normal AGN.  Due to the high redshift of PSO J334 standard emission line diagnostics, including narrow emission line ratios (e.g., \citealt{Kewley2006}) or broad H$\alpha$ emission (e.g., \citealt{Greene&Ho2005}), are redshifted into the near-infrared where archival spectroscopic data are not available.  However, the good agreement between PSO J334's SED and the non-blazar AGN SEDs presented in \cite{Shang2011} strongly suggests a single-AGN system. %

We compute the IR colors of PSO J334, as IR colors are often used as a tool to identify AGN \citep{Jarrett2011,Stern2012}, although red IR colors can also be produced when dust is heated by recent star formation.  Specifically, we determine the position of PSO J334 on a \emph{WISE} color--color plot to (i) determine whether it meets the AGN IR criteria, and (ii) determine if there are any abnormalities in its placement with respect to a normal AGN.  We compare PSO J334's \emph{WISE} colors to the ``AGN box" empirically defined by \cite{Jarrett2011}, which is based on the colors of quasi stellar objects (QSOs) and Seyfert galaxies with redshift out to $z \sim 2$.  With $W1 - W2 = 1.13 \pm 0.11$ and $W2 - W3 = 3.26 \pm 0.27$, PSO J334 has IR colors located within the AGN box of \cite{Jarrett2011} and does not seem to have any IR color abnormalities compared to other QSOs at similar redshifts.  However, it is not clear if binary-AGN would differ from a standard AGN in IR color-color diagrams (see \citealt{Ellison2017} for an example where a candidate dual-AGN has \emph{WISE} colors consistent with a standard AGN).

Secondly, our \emph{Chandra} observations allows us to look at the location of PSO J334 on the Fundamental Plane of Black Hole Activity (FP)\,---\,an empirical relationship between the black hole mass, 5 GHz luminosity $L_{5GHz}$, and 2--10 keV luminosity $L_{X}$.  \cite{Gultekin2009a} fit a relation to be used as an estimation for black hole mass based on observations of $L_{X}$ and $L_{5GHz}$ of the form:
\begin{equation}
\resizebox{0.5\textwidth}{!}{log$\Big(\frac{M}{10^{8}M\scriptscriptstyle\odot}\big) = 0.19 +  0.48$log$\Big(\frac{L_{5\mathrm{GHz}}}{10^{38}\,\mathrm{erg\,s^{-1}}}\Big) - 0.24$log$\Big(\frac{L_{X}}{10^{40}\,\mathrm{erg\,s^{-1}}}\Big).$}   
\end{equation}
Whether  or not the local FP relation is appropriate for high accretion rate AGNs remains a topic of debate.  For example, \cite{Gultekin2009a} studied a sample of AGN with dynamical black hole mass estimates and uniform analysis of archival \emph{Chandra} data, and confirmed that the inclusion of high-accretion rates sources, such as Seyfert galaxies, increased the intrinsic scatter about the FP. However, other analyses reflect that the FP may be applicable for high accretion-rate sources (e.g., \citealt{Panessa2007,Gultekin2014}).  Recognizing the caveats of such an analysis, we use the results of \cite{Gultekin2009a} to calculate the projected mass of the system and compare it to the measured total mass of the system, which we take to be log$(M/M_{\odot})=9.1$ \citep{liu2016,mooley2017}.  Binary-AGN with small mass ratios may result in offsets from the FP if the secondary is the main accretor\,---\,the mass calculated via the FP relation may be smaller compared to mass of the entire system, reflecting that the coupled X-ray/Radio emission stems from the less massive secondary. However, because of the large scatter on the calculated mass via the FP relation of $\sim1\,\mathrm{dex}$, this approach is only sensitive to the more extreme mass ratio values, e.g. $q \lta 0.01$.  We find that the mass projected from the FP relation of \cite{Gultekin2009a} is consistent, within the error bars, with the entire mass of the system. 

\subsection{Binary-AGN models}
\subsubsection{A Cavity in the Circum-binary Disk}
If PSO J334 were consistent with the cavity model, the radiation that a normal disk would radiate within the inner edge of the circum-binary disk, $R_\mathrm{circum,edge}$, will be missing.  The exact values of the temperatures we deduce below depend on details regarding the micro-physics of the system; here we make rough estimates using a standard accretion disk model.  Assuming a thin-disk model, a mass ratio not much less than unity (e.g., $q > 0.01$), and an inner edge located at $R_\mathrm{circum,edge} \simeq 2a$ \citep{Milosavljevic2005}, the temperature at the inner edge of the circum-binary disk is given by:
$T_\mathrm{circum,edge} \simeq 1.96 \times 10^{4} \big[ \dot{m}\big(\eta/0.1\big)^{-1}M_{8}^{-1}\big(a/100R_{G}\big)^{-3}\big]^{1/4}$ K \citep{Roedig2014}. Here, $\dot{m}$ is the accretion rate in Eddington units, $\eta$ is the accretion efficiency, $M_{8}$ is the total mass of the binary in units of $10^{8} M_{\odot}$, and $a$ is the separation between the two BHs in units of gravitational radius $R_{G}$ ($R_{G} = GM/c^{2}$). In the case of a cavity, we do not expect much emission at temperatures above $T_\mathrm{circum,edge}$. We note that this temperature is not predicted to change late into the evolution of the binary where the orbital time becomes smaller than the inflow time of the circum-binary disk, a scenario that may be very relevant to PSO J334 given the calculated separation between the SMBHs listed in \cite{liu2016}.  

To calculate $T_\mathrm{circum,edge}$ we consider the parameters $\eta = 0.1$, log$(M/M_{\odot})=9.1 \pm 0.3$ \citep{mooley2017}, and $a = 28 \pm 14R_{S} = 56 \pm 28R_{G}$ \citep{liu2016}. We calculate the accretion rate in Eddington units as $\dot{m} = {\dot{M}} / {\dot{M}_\mathrm{Edd}} \approx {\dot{M}}[{3\times10^{-8} (M/M_{\scriptscriptstyle \odot}) M_{\scriptscriptstyle \odot} \textrm{yr}^{-1}}]^{-1}$  $\approx 0.26 \pm 0.1$.  Here $\dot{M} = {L_\mathrm{bol}}/ ({\eta c^{2}})$, where $L_\mathrm{bol}$ is the bolometric luminosity of the circum-binary accretion disk and is determined from the quasar bolometric luminosity corrections presented in \cite{Runnoe2012}, using the quasar's continuum flux density at $\lambda =3000$ \angstrom ~presented in \cite{liu2016}.  

$T_\mathrm{circum,edge}$ is found to be $\simeq 11000 \pm 4000$ K, or a cut-off wavelength $\lambda_\mathrm{circum,edge} \simeq 2500^{+1600}_{-700}$ \angstrom ~assuming blackbody radiation from the circum-binary accretion disk.  We indicate the position of $\lambda_\mathrm{circum,edge}$ on PSO J334's SED in Fig~\ref{fig:2}.  If PSO J334 were consistent with the cavity accretion model, we would expect emission with energies higher than $\sim$NUV to be either (i) significantly lower than expected from a normal AGN disk or (ii)  disappear entirely.    However, if PSO J334 were consistent with a single-AGN system, we would expect UV emission from the inner-most part of the accretion disk that upscatters to X-rays via inverse-Compton interactions with the corona.
Combining our 2--10 keV detection and the \emph{GALEX} archival data, we can verify that the X-ray intensity is consistent with the expected upscattered UV emission from a normal AGN disk, reflected by the considerable overlap between PSO J334's SED and the \cite{Shang2011} data in Fig.~\ref{fig:2}. We conclude that PSO J334 is incompatible with the cavity scenario.

\subsubsection{``Notches" from a Minidisk}
Regarding the minidisk model, the radiation that an ordinary disk would radiate between the inner edge of the circum-binary disk and the tidal truncation radii of the minidisks, $R_\mathrm{tidal}$, will be missing.  In such a scenario, it is expected that the missing emission will produce a dip, or a ``notch", in the thermal continuum spectrum, reflecting the missing emission between $R_\mathrm{circum,edge}$ and $R_\mathrm{tidal}$ (e.g. \citealt{Roedig2012,Tanaka2012,Gultekin&Miller2012,Kocsis2012, Tanaka&Haiman2013, Roedig2014}, however see \citealt{Farris2015a} for a simulation where notches become obscured).  Thus, it is possible to use PSO J334's SED to search for evidence of minidisks.   Specifically, \cite{Roedig2014} derive analytical calculations of the specific luminosity integrated from the cirum-binary disk and the two mini-disks, where the primary and secondary BHs are accreting material at rates $\dot{M_{1}} = f_{1}\dot{M}$ and $\dot{M_{2}} = f_{2}\dot{M}$. Here $\dot{M_{1}}$ and $\dot{M_{2}}$ are the mass accretion rates of the primary and secondary, and $\dot{M}$ is the circum-binary accretion rate.  Further, they assume the circumbinary disk is in inflow equilibrium (i.e., $f_{1}+f_{2}=1$) and take into account a hardening factor $g=1.7$. The emergent spectrum may be hardened by a factor $g$ due to the majority of the disk laying in a regime in which electron scattering opacity dominates absorption opacity \citep{Shimura&Takahara1995}.  Lastly, they assume that there is a sharp surface density cut-off at the inner edge of the circumbinary disk and the outer edges of the mini-disks.  For this particular model, they find that a spectral depression tends to occur in the SED between $\sim kT_\mathrm{notch} - 15 kT_\mathrm{notch}$, where $T_\mathrm{notch}$ is the characteristic temperature of the accretion disk at a radius $r_\mathrm{notch}\sim a$ and is approximately $2^{3/4}T_\mathrm{circum,edge}$ (see their equation 3). $T_\mathrm{notch}$ is evaluated at a radius that lies between the hottest point in the circum-binary accretion disk (which is truncated to $\sim 2a$) and the coldest point in the minidisk (which extends to $\sim a/2$). We expect very little thermal radiation at the energy corresponding to the deepest point of the notch, which is centered at $E\simeq 4kT_\mathrm{notch}$. 

For PSO J334, we calculate $T_{notch} \approx 19000 \pm 7000$ K, translating to an observable notch in the SED between 500--7000 \angstrom, where the deepest point is predicted to occur at $\lambda_\mathrm{notch} \approx 1900^{+1200}_{-500}$ \angstrom.  In Fig.~\ref{fig:2} we indicate the position of $\lambda_\mathrm{notch}$ with respect to PSO J334's SED.  We use the analytical calculations in \cite{Roedig2014} to analyze how the SED shape is affected by the presence of a notch, for various parameters of $q$, $f_{1}$, and $f_{2}$.   The continuum for the notch is estimated by approximating the \cite{Shang2011} SED between 500 \angstrom ~and 7000 \angstrom~with a straight line.  Three examples of notched SEDs are shown in Fig.~\ref{fig:2}.  Our SED is reasonably resolved throughout the expected frequency range of a spectral notch, and the data appear nominally closer to the standard \cite{Shang2011} SED model.

We note that it is possible the mass ratio of PSO J334 falls much below $q \simeq 0.3$ \citep{liu2015}.  Specifically, for the case of $q \le 0.1$ it is likely that any notch in the SED will have a different shape from the analysis above.  We may expect $\lambda_\mathrm{notch}$ to occur at shorter wavelengths and the deepest portion of the notch to be even lower; this is a result of the primary BH's accretion flow barely contributing to the total SED (i.e., $f_2 \gg f_1$; \citealt{Roedig2014}).  High S/N spectroscopy between 500 \angstrom ~and 7000 \angstrom, along with dense FUV observations, are required for a more robust analysis on possible notches in the SED.

\subsubsection{Hard X-ray Emission from Minidisks}
As a result of the expected supersonic motions of streams that are accreting onto the minidisks from the circum-binary disk, shocks are predicted as the streams hit the minidisk edges.  It has been shown that these shocks should manifest in an excess of hard X-ray emission in the SED \citep{Roedig2014,Farris2015a,Farris2015b}.  \cite{Roedig2014} show that the post-shock temperature, $T_\mathrm{{ps}}$, of accreting streams is usually in excess of $10^{9}$ K, with $T_\mathrm{{ps}} \propto (a/100R_{G})^{-1}(1+q^{0.7})^{-1}$ (assuming the secondary is the main accretor).  However, cooling is expected to rapidly set in such that the resultant emission is between 50 and 200 keV.  These results agree with simulations presented in \cite{Farris2015b}, where excess emission from stream shocking is $\sim$10 times higher than a normal AGN SED between 10 and 100 keV. Given that \cite{Farris2015b} assume a mass ratio $q = 1.0$, PSO J334 may be expected to have an excess of emission at an even energy higher if the mass ratio of the binary-AGN is much lower than unity. %

Our rest-frame $\sim$ 0.9 to 24.5 keV \emph{Chandra} spectrum shows no evidence of excess hard X-ray emission with respect to an absorbed power-law model.  However, as argued above, it is likely that any excess X-ray emission would reside above our \emph{Chandra} observation's energy range.  We consider possible temperatures for the stream-shocking emission from PSO J334, given an assumed semi-major axis $a=56R_{G}$ and a range of mass ratios $0.01 < q < 1.0$.  Adopting the assumption from \cite{Roedig2014} that a mass ratio $q \simeq 1.0$ and semi-major axis $a=100R_{G}$ will result in an additional Wien-like spectrum with peak energy $\simeq 100$ keV, we predict that any excess emission from stream-shocking in PSO J334 could peak between rest-frame 180--340 keV, or observed-frame 60--110 keV.  Because of the approximate nature of these calculations, we can only make an estimate for detectability with \emph{NuSTAR}.  However, if we assume a similar photon index in the 60--110 keV energy range as found in our 2--10 keV spectral fit, and assume that stream-shocking will result in an excess of emission $\sim$10 times higher than a normal AGN SED in this energy range, the count rate in \emph{NuSTAR} is expected to be close to $\sim$1$\times$10$^{-5}$ counts per second.  Thus, such emission will not be easily detected by \emph{NuSTAR}.  We note that a more detailed analysis on the post-shock temperature evolution, including relevant cooling mechanisms such as pair production, will be necessary in order to determine the resultant hard X-ray spectrum expected from stream shocking.

\section{Conclusions}
In this work, we present the first targeted X-ray observation of PSO J334, a candidate binary-AGN system, with the aim to uncover the true accretion nature of the quasar. If a binary-AGN system  with a separation of 28$R_{S}$ \citep{liu2016}, PSO J334 is well into the gravitational-wave dominated regime and should be undergoing circum-binary accretion. Simulations show that two main types of circum-binary accretion disk morphologies can be expected: a “cavity”,  where emission is truncated blueward of the wavelength associated with the temperature of the innermost ring due to a mostly empty accretion disk, or minidisks, where there is substantial accretion onto one or both of the members of the binary, each with their own shock-heated thin disk. Cavities and minidisks are expected to exhibit different behavior in the high-energy regime.  Specifically, if the accretion disk of PSO J334 contains a cavity it is very likely that no, or very little, X-ray emission is expected. Further, if PSO J334 is accreting via minidisks, we may expect to see a notch in the SED or an excess of hard X-ray emission above $\ge 100$ keV.  Our 40 ks \emph{Chandra} observation allows for the opportunity to discern between a single- or binary-AGN system, and if a binary, can characterize the type of circum-binary accretion.  The main results and implications of this work can be summarized as follows: \\
(1) We find that PSO J334's X-ray emission is best explained by a mildly absorbed power-law with intrinsic  $N_{H} = 0.91^{+4.84}_{-0.89} \times 10^{22}$ cm$^{-2}$  and $\Gamma = 2.02^{+0.83}_{-0.39}$, with an observed 2--10 keV flux of $3.20^{+0.9}_{-1.1} \times$ 10$^{-14}$ erg cm$^{-2}$ s$^{-1}$, or rest-frame 2--10 keV luminosity of $9.40^{+1.4}_{-1.1} \times$ 10$^{44}$ erg s$^{-1}$ at $z=2.06$ (assuming isotropic emission).  We fit a broken power-law model, which may originate from a binary system where both SMBHs are accreting with their own minidisk, and find at a 95\% confidence level that the spectrum does not need an additional photon index to explain its shape. \\
(2) We construct a radio--X-ray SED for PSO J334, using new VLBA data \citep{mooley2017}; archival \emph{WISE}, UKIDSS, Pan-STARRS1 \citep{liu2015}, and \emph{GALEX} data; and our new \emph{Chandra} data. We find the SED agrees well with the composite non-blazar AGN SEDs presented in \cite{Shang2011}.  \\
(3)  Other analyses, such as comparing IR \emph{WISE} colors to the empirical ``AGN box" presented in \cite{Jarrett2011} and calculating the mass of the accreting system via the FP, further reflects the similarity of PSO J334 with respect to normal AGN. \\
(4) We calculate the temperature at the inner edge of a possible circum-binary disk and find that no, or very little, emission is expected beyond $\sim 2500^{+1600}_{-700}$ \angstrom.  However from our \emph{Chandra} observation we can verify that the intensity at energies above 2500 \angstrom ~is consistent with the expected upscattered UV emission from a normal AGN disk. We conclude that the X-ray emission of PSO J334 is incompatible with the cavity accretion mode. \\
(5)  We find no gap in the SED expected from the missing emission between $R_\mathrm{circum, edge}$ and $R_\mathrm{tidal}$, predicted to be between 500--7000 \AA, for mini-disk accretion models. We note that it is possible that a notch exists within the data but is undetected given the resolution of our SED. \\
(6) If PSO J334 is accreting via minidisks, then we may expect a detectable excess of hard X-ray emission above $E \ge 100$ keV, depending on the mass ratio of the system and the various time-dependent cooling processes of the post-shock photons.  Our rest-frame $\sim$0.9 to 24.5 keV \emph{Chandra} spectrum shows no  evidence  of  excess  hard  X-ray  emission  with  respect  to an absorbed power-law model, however we  predict  that  any  excess emission from stream-shocking should peak between rest-frame 180--340 keV, or observed-frame 60--110 keV.  Near-future observations are unlikely to detect a possible excess of emission at these higher energies. \\

We have shown through various analyses that there is an absence of evidence supporting PSO J334 as a binary-AGN system.  Specifically, we find no compelling evidence supporting PSO J334 as a binary-AGN system containing a cavity, or a binary-AGN system with mass ratios $q\ge0.1$.  However, because of the small number of currently promising binary-AGN candidates, it is most likely that the best method to distinguish a binary-AGN from a single-AGN has yet to be identified.  Regarding the true nature of PSO J334, a stronger argument in either direction can be made with i) a high S/N spectrum between 500--7000 \AA~to allow for a more robust analysis on whether PSO J334 agrees better with the standard AGN-model or a notched-SED, and ii ) a FUV spectrum that will allow for a better analysis of a possible binary-AGN system with mass ratios much below $q=0.1$. As well, hard X-ray observations targeting excess emission expected from stream-shocking will be important for determining the accretion mode of the quasar.

\hypertarget{ackbkmk}{}%
\acknowledgements 
\bookmark[level=0,dest=ackbkmk]{Acknowledgments}
We thank the referee for helpful and constructive comments.  AF acknowledges support provided by the Rackham Graduate School Conference Grant and the National Space Grant Foundation's John Mather Nobel Scholarship Program. KG and MR acknowledge support provided by the National Aeronautics and Space Administration through Chandra Award Number GO6-17104X issued by the Chandra X-ray Observatory Center, which is operated by the Smithsonian Astrophysical Observatory for and on behalf of the National Aeronautics Space Administration under contract NAS8-03060.
This research has made use of the NASA/IPAC Extragalactic Database
(NED) which is operated by the Jet Propulsion Laboratory, California
Institute of Technology, under contract with the National Aeronautics
and Space Administration.  
This research has made use of NASA's Astrophysics Data System.

\bibliographystyle{apjads}
\hypertarget{refbkmk}{}%
\bookmark[level=0,dest=refbkmk]{References}
\bibliography{foord.bib}

\label{lastpage}
\end{document}